\documentstyle[preprint,aps,epsf]{revtex}

\newcommand{\Tr}{{\rm Tr}}

\tightenlines

\begin{document}

\draft

\title{
\vspace{-4.0cm}
\begin{flushright}
\normalsize
ITP-Budapest 547\\
UTCCP-P-60\\
UTHEP-397\\
hep-lat/9901021\\
\vspace{2pt}
Jan 1999
\end{flushright}
The endpoint of the first-order phase transition
of the SU(2) gauge-Higgs model\\
on a 4-dimensional isotropic lattice}

\author{
  Y.~Aoki$^{1, 2}$,
  F.~Csikor$^{3}$,
  Z.~Fodor$^{3}$, and
  A.~Ukawa$^{1, 2}$
}

\address{
  ${}^{1}$Center for Computational Physics, University of Tsukuba,
  Tsukuba, Ibaraki 305-8577, Japan\\
  ${}^{2}$Institute of Physics, University of Tsukuba,
  Tsukuba, Ibaraki 305-8571, Japan\\
  ${}^{3}$Institute for Theoretical Physics, E\"otv\"os University,
  H-1088 Budapest, Hungary\\
}

\maketitle

\begin{abstract}

We study the first-order finite-temperature electroweak 
phase transition of the SU(2) gauge-Higgs model defined on a 
4-dimensional isotropic lattice with
temporal extension $N_t=2$.
Finite-size scaling study of Lee-Yang zeros yields the value of the
Higgs self coupling of the endpoint at $\lambda_c=0.00116(16)$.
An independent analysis of Binder cumulant gives 
a consistent value for the endpoint. 
Combined with our zero-temperature measurement of Higgs 
and $W$ boson masses, this 
leads to $M_{H,c}=73.3\pm 6.4$ GeV for the critical Higgs boson mass 
beyond which the electroweak transition turns into a crossover. 
\end{abstract}


\newpage

\section{Introduction}

The Minimal Standard Model predicts that the electroweak interaction 
undergoes a first-order phase transition at a finite temperature for 
light Higgs boson masses.
A focus of recent studies has been whether the first-order phase transition
survives with sufficient strength for a realistically heavy Higgs boson 
mass \cite{RummLatt96}, 
since the feasibility of electroweak baryogenesis  
 \cite{Rubakov96} depends crucially on it.

The first-order nature of the electroweak transition for light Higgs 
bosons can be shown within perturbation theory.  However, 
perturbation theory breaks down for Higgs boson masses larger than about 
$M_W$ due to bad infrared behavior of the gauge-Higgs part of 
the electroweak theory \cite{Fodor}.  Hence numerical simulation 
techniques are needed to analyze the nature of the transition 
for heavy Higgs bosons. 

Extensive studies in this direction have already been performed 
within the effective 3-dimensional theory approach, 
in which all non-static modes of the system are integrated out
 perturbatively.  This approach has the advantage 
that the full Standard Model including fermions can be mapped onto 
a 3-dimensional SU(2) (or $\mbox{SU(2)}\otimes \mbox{U(1)}$) 
gauge-Higgs model, as there are no fermionic static modes
at finite temperature.  In addition, thinning out the degree of 
freedom to those of a 3-dimensional theory significantly reduces 
the computational requirement. 

Results from simulations in this approach show that the first-order 
electroweak transition weakens as the Higgs boson mass increases
\cite{Kajantie1,Leipzig96,Karsch96}, 
and that it turns into a continuous crossover for heavy Higgs bosons 
with a mass $M_H \gtrsim M_W$\cite{Kajantie2}. 
Detailed studies of the endpoint of the 
first-order transition including its universality class 
have also been made\cite{KarschLat96,Leipzig97,Rumm98}. 
   
A potential problem with the 3-dimensional approach is that it relies
on perturbation theory to derive the 3-dimensional action so that numerical 
predictions 
may involve systematic errors due to truncation of perturbative series. 
From this point of view a direct simulation of the 4-dimensional system 
is preferred. Results from 4-dimensional simulations provide a check on 
those of the 3-dimensional method.  

Early studies of the 4-dimensional SU(2) gauge-Higgs system were carried out
in Refs.~\cite{Leipzig93,DESY94,DESY95,DESY96,yaoki}. More recently 
advances have been made with the use of space-time anisotropic 
lattice \cite{Csikor981,Csikor982}.  This approach alleviates
the double-scale problem that there are light modes 
with long wave length, $\xi >> 1/T$, near the endpoint where the transition 
is of second order. 

In this article we report on a study of the endpoint of the SU(2) gauge-Higgs 
model employing 4-dimensional space-time symmetric lattices with the 
temporal lattice size $N_t=2$, building 
upon a previous work \cite{yaoki}.  Simulations have been carried out 
for a wide range of spatial lattice sizes, and finite-size scaling study of 
Lee-Yang zeros is used to find the location of the endpoint.  We measure 
the Higgs and W boson masses around the endpoint
and estimate the value of the Higgs boson mass at the endpoint. 

This paper is organized as follows. 
In Section \ref{sec:theory} we present the SU(2) gauge-Higgs model lattice 
action and outline our strategy for finding the endpoint through Lee-Yang
zeros. 
In Section \ref{sec:finiteT}, following a brief discussion of 
susceptibility analysis, Lee-Yang zeros are examined. 
Another approach to find the endpoint using the Binder cumulant is also 
described. 
In Section \ref{sec:zeroT} we present results of the zero-temperature
mass measurement.  Together with our result for the scalar 
self-coupling constant at the endpoint obtained through Lee-Yang zero 
analysis, this leads to the value of the Higgs boson mass 
at the endpoint.
Sec.~\ref{sec:conc} is devoted to conclusions.

\section{Theory and Simulation}
\label{sec:theory}

We work with the standard SU(2) gauge-Higgs model action given by 
\begin{equation}
\label{eq:S_lat}
  S =  \sum_x \left[
    \sum_{\mu > \nu} \frac{\beta}{2} \Tr U_{x,\mu \nu}
    \mbox{} + \mathop{\sum}_\mu 2 \kappa L_{x, \mu}
    \mbox{} - \rho_x^2 - \lambda ( \rho_x^2 -1 )^2
    \right],
\end{equation}
\begin{equation}
  \label{eq:link}
    L_{x, \mu}\equiv \frac{1}{2}
    \Tr ( \Phi_x^{\dag} U_{x, \mu} \Phi_{x+\hat{\mu}} ),
    \  \rho_x^2\equiv\frac{1}{2}\Tr (\Phi_x^{\dag}\Phi_x),
\end{equation}
where $U_{x,\mu \nu}$ is the product of link operators around a plaquette, 
$\beta$ is related to the tree-level gauge coupling as $\beta=4/g^2$,  
$\kappa$ represents the Higgs field hopping parameter and 
$\lambda$ is the scalar self-coupling.
We put the system on a space-time isotropic lattice of a size 
$N_t \times N_s^3$.

Finding the endpoint of the first-order 
finite-temperature phase transition of the model 
requires finite-size scaling analyses to quantitatively distinguish 
the case of a first-order transition from that of a crossover as the coupling 
parameters of the model are varied.  As the main tool, we employ 
finite-size scaling analysis of Lee-Yang zeros \cite{LeeYang,Itzykson}
on the complex $\kappa$ plane for fixed $\beta$ and $\lambda$ 
\cite{KarschLat96,Leipzig97,Csikor982}.
For a first-order 
phase transition, the infinite volume limit of the zeros pinches the real 
$\kappa$ axis, while they stay away from it if there is no phase transition. 
We also supplement this method with analyses of susceptibility and Binder 
cumulant.

Our finite-temperature simulations are carried out for the temporal 
lattice size $N_t=2$.  For the spatial lattice size we take 
$N_s^3=20^3,24^3,32^3,40^3,50^3$ and $60^3$.
The gauge coupling is fixed at $\beta=8$. 
For the scalar self-coupling we choose  five values, 
$\lambda=0.00075, 0.001, 0.00135, 0.00145$ and $0.0017235$,
which covers the range of zero-temperature Higgs boson mass
$57 \lesssim M_H \lesssim 85$GeV \cite{yaoki}.
For each value of $\lambda$ the scalar hopping parameter $\kappa$ is 
tuned to the vicinity of the pseudo critical point estimated by 
the peak position of the susceptibility of the Higgs field length 
squared $\rho^ 2$. 

The updating algorithm is a combination of over-relaxation and heatbath 
methods \cite{DESY95}, with the ratio of the two for the scalar part and the 
gauge part as specified in Ref.~\cite{yaoki}.
We make at least $10^5$ iterations of this hybrid over-relaxation
algorithm at each coupling parameter point for each lattice size.
The list of coupling values and statistics we use in our finite-temperature 
simulations are listed in Table \ref{tab:paramT}.

We also carry out zero-temperature simulations to measure the masses of 
Higgs and W bosons around the endpoint of the first-order phase transition. 
For these runs an improved algorithm of Ref.~\cite{BunkLatt94} is employed. 
Details of the runs and results are discussed in Sec.~\ref{sec:zeroT}.

\section{Finite-temperature Results}
\label{sec:finiteT}

\subsection{Susceptibility}

\begin{figure}[p]
  \begin{center}
    \leavevmode
    \epsfxsize=10cm \epsfbox{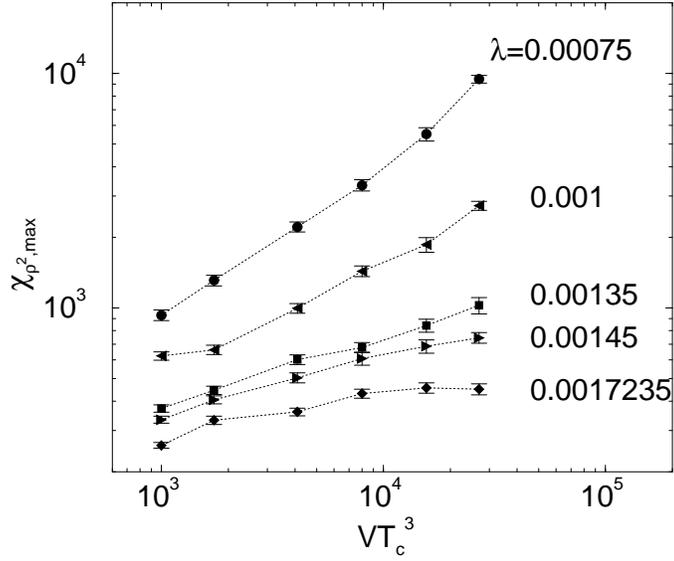}
    \caption{Peak height of susceptibility of $\rho^2$ against inverse volume
      normalized by critical temperature $VT_c^3=N_s^3/N_t^3$.
      Dotted lines are guides for eyes.}
    \label{fig:sus}
  \end{center}
\end{figure}

Let us first look at the susceptibility of squared Higgs length,
\begin{equation}
  \label{eq:sus_rho}
  \chi_{\rho^2} \equiv V \left( \langle \rho^2 \rangle
    \mbox{} - \langle \rho \rangle^2 \right),
\end{equation}
where $V\equiv N_s^3$. 
The maximum value of the susceptibility at its peak, calculated by the 
standard reweighting technique \cite{reweighting} as a function of $\kappa$, 
is plotted in Figure \ref{fig:sus} against the spatial volume normalized 
by the critical temperature $VT_c^3=N_s^3/N_t^3$.  Errors are estimated 
by the jackknife procedure with the bin size of $10^3$--$10^4$ sweeps,
which is listed in Table \ref{tab:paramT}.

The slope for the smallest scalar coupling $\lambda=0.00075$ approaches 
unity for large volumes, which is consistent with a first-order transition, 
while that for the largest coupling $\lambda=0.0017235$ tends to 
a constant, showing an absence of a phase transition. 
A continuous decrease of the slope for the intermediate values of $\lambda$ 
indicates that the endpoint of the first-order transition is located 
in between the two extreme values.  Our range of spatial volumes, 
unfortunately, is not sufficient to pin down the critical value of $\lambda$ 
from the susceptibility data. 

\subsection{Lee-Yang Zeros}
\label{subsec:ly0}

The determination of the endpoint of the finite temperature
phase transition of the model, thus a characteristic feature of the phase 
diagram,
is made by the use of the Lee-Yang zeros of the
partition function $Z$ \cite{LeeYang,Itzykson}.
Near the first-order phase transition point the partition function reads
\begin{eqnarray}
 Z= Z_s +  Z_b \propto \exp (-V f_s ) + \exp ( -V f_b ) \, ,
\end{eqnarray}
where the indices s(b) refer to the symmetric (Higgs) phase and $f$ stands
for the free-energy densities. Near the phase transition point we also have
\begin{eqnarray}
f_b = f_s + \alpha (\kappa - \kappa _c ) \, ,
\end{eqnarray}
\noindent
since the free-energy density is continuous. One then obtains 
\begin{eqnarray}
 Z\propto \exp [ -V ( f_s +f_b )/2 ] \cosh [ -V \alpha (\kappa -\kappa_c )/2] \,
\
\end{eqnarray}
which shows that for complex $\kappa$ $Z$ vanishes at
\begin{eqnarray}\label{eq:zeros}
{\rm Im} (\kappa )= 2\pi \cdot (n-1/2) / (V\alpha )
\end{eqnarray}
for integer $n$.  In case a first-order phase transition is present,
these Lee-Yang
zeros move to the real axis as the volume goes to infinity. If a 
phase transition is absent the Lee-Yang
zeros stay away from the real $\kappa $ axis. Thus the way the Lee-Yang
zeros move in this limit is a good indicator for the presence or
absence of a first-order phase transition.

Calculation of the partition function for complex values of $\kappa$ is 
made with the reweighting method \cite{reweighting}
in both imaginary and real directions of $\kappa$. 
In those cases where we have two ensembles with the same value of $\lambda$ 
and $N_s$, but different $\kappa$, we combine the two runs by setting 
the magnitude of the two partition functions
to be equal at the midpoint between the two $\kappa$'s.

\begin{figure}[p]
  \begin{center}
    \leavevmode
    \epsfxsize=10cm  \epsfbox{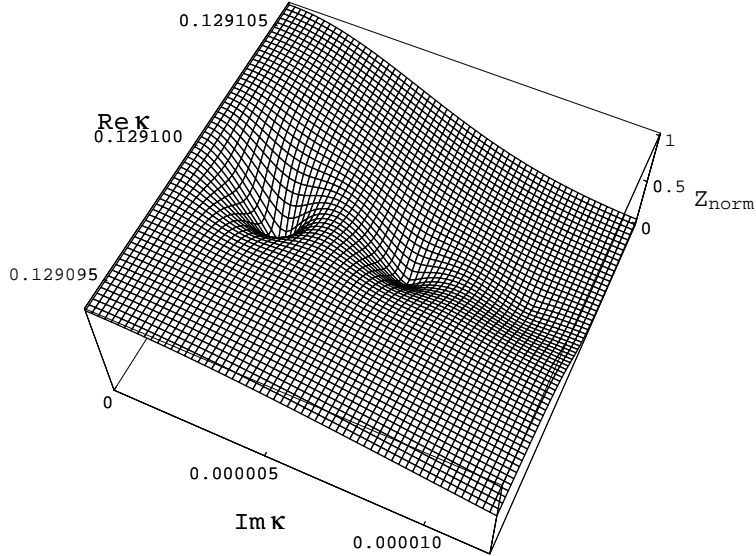}
    \caption{Absolute value of normalized partition function as a function
      of complex $\kappa$ for $\lambda=0.00075$ and $N_s=60$.}
    \label{fig:abs_z_f60d}
  \end{center}
\end{figure}

\begin{figure}[p]
  \begin{center}
    \leavevmode
    \epsfxsize=8cm \epsfbox{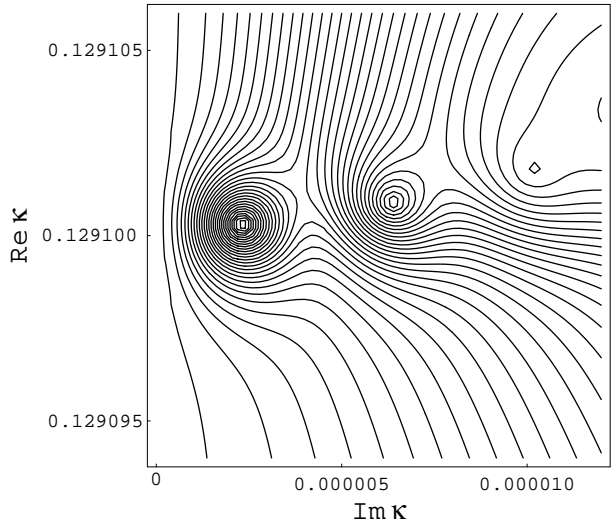}
    \caption{Contour plot of Fig.~\protect\ref{fig:abs_z_f60d}.}
    \label{fig:z_cont_f60d}
  \end{center}
\end{figure}

In Fig.~\ref{fig:abs_z_f60d} we show the absolute value
of the partition function normalized by its value at the real axis 
on the complex $\kappa$ plane,
\begin{equation}
  \label{eq:z_norm}
   Z_{norm}(\kappa) \equiv
  \left| \frac{Z({\rm Re}\kappa,{\rm Im}\kappa)}{Z({\rm Re}\kappa,0)} \right|
\end{equation}
for $\lambda=0.00075$ and $N_s=60$.
The contour line of this figure is shown in Figure \ref{fig:z_cont_f60d}.
We observe three zeros in this case, whose distance from the real axis 
is roughly in the ratio $1:3:5$ as expected from (\ref{eq:zeros}) 
for a first-order transition. 

\begin{figure}[p]
  \begin{center}
    \leavevmode
    \epsfxsize=10cm \epsfbox{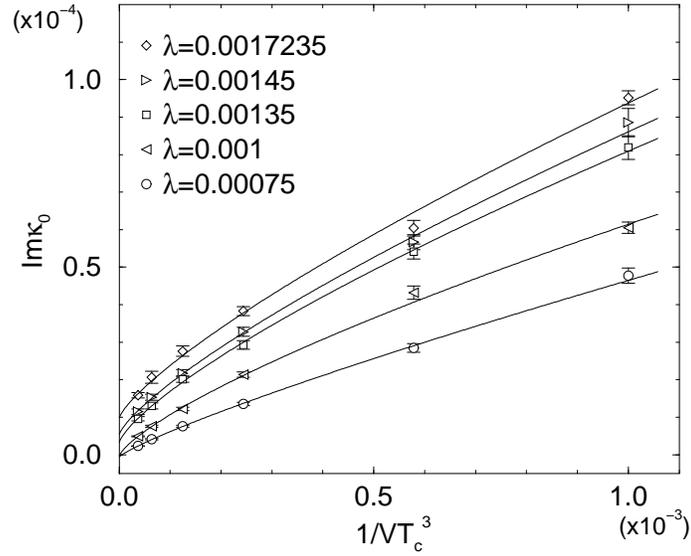}
    \caption{Imaginary part of first Lee-Yang zero as a function of inverse
      volume normalized by the critical temperature.
      Solid lines are least $\chi^2$ fits with 
      ${\rm Im}\kappa_0(V) = \kappa_0^c + C V^{-\nu}$.}
    \label{fig:fit_ly0}
  \end{center}
\end{figure}

Let us call the zero nearest to the real axis as first zero, and denote 
its location by $\kappa_0$. 
We search for the first zero by the Newton-Raphson method applied to 
the equation
\begin{equation}
  \label{eq:ly0}
   Z({\rm Re} \kappa, {\rm Im} \kappa) = 0,
\end{equation}
starting with an initial guess for $\kappa_0$ obtained from the contour 
plot of $Z_{norm}(\kappa)$.  
The error of $\kappa_0$ is estimated by the jackknife method with a bin 
size given in Table \ref{tab:paramT}, {\it i.e.,} the zero search is repeated 
for the set of partition functions calculated from each jackknife sample of 
configurations, and the jackknife formula is applied to the set of $\kappa_0$.
The results for $\kappa_0$ are given in Table \ref{tab:paramT}.
We show in Fig.~\ref{fig:fit_ly0}
values of the imaginary
part of the first zero ${\rm Im}\kappa_0(V)$ as a function of inverse volume. 

Finite-size scaling theory predicts that the volume dependence of 
the imaginary part of the first zero is given by a scaling form,  
\begin{equation}
 \label{eq:fss_ly0}
   {\rm Im}\kappa_0(V) = \kappa_0^c + C V^{-\nu}. 
\end{equation}
For a first-order phase transition, the infinite volume limit vanishes, 
$\kappa_0^c=0$, and the exponent takes the value $\nu=1$.  In the absence of 
a phase transition, $\kappa_0^c\ne 0$ and the value of the exponent is 
generally unknown.  

\begin{figure}[p]
  \begin{center}
    \leavevmode
    \epsfxsize=10cm \epsfbox{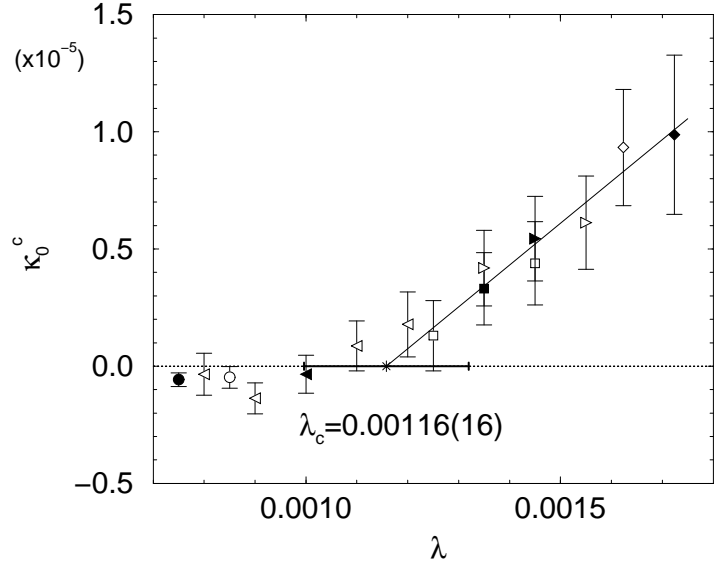}
    \caption{Imaginary part of first Lee-Yang zero at infinite-volume limit
      as a function of Higgs self coupling. 
      Filled symbols are calculated without $\lambda$-reweighting,
      while open symbols with
      $\lambda$-reweighting from the filled symbol with same shape.
      Solid line is a linear fit to
      $\lambda=0.00135, 0.00145$ and $0.0017235$ (filled symbols).}
    \label{fig:res_ly0}
  \end{center}
\end{figure}

In Fig.~\ref{fig:res_ly0} we plot results for $\kappa_0^c$ as a function of 
$\lambda$ obtained by fitting the volume dependence of the first zero by 
the form (\ref{eq:fss_ly0}) (see Fig.~\ref{fig:fit_ly0} for fit lines).  
Both $\kappa_0^c$ and $\nu$ are taken as fit 
parameters, and the entire set of volume $N_s^3=20^3-60^3$ is employed.  
Filled symbols mean that they are directly obtained from the 
simulations carried out at the corresponding values of $\lambda$. 
The points plotted with open symbols are obtained from the 
first zero of the partition function calculated by reweighting 
the partition function measured at the point where 
$\kappa_0^c$ with the filled symbol of the same shape is shown. 
The agreement of open symbols of different shapes within errors 
shows that reweighting from different values of $\lambda$ gives 
consistent results between the measured points.

At small couplings $\lambda \lesssim 0.001$, $\kappa_0^c$ is consistent
with zero, which agrees with the result of Ref.~\cite{yaoki}
that the transition is of first order in this region.
At large couplings $\lambda \gtrsim 0.0013$, $\kappa_0^c$ no longer
vanishes, and hence there is no phase transition.
In order to determine the endpoint of the phase transition, 
we take the three filled points at $\lambda=0.00135, 0.00145$ and $0.0017235$
directly obtained from independent simulations without $\lambda$-reweighting, 
and make a fit with a function linear in $\lambda$.
This gives the position of the endpoint to be 
\begin{equation}
\label{eq:endpoint}
\lambda_c = 0.00116(16).
\end{equation}

\begin{figure}[p]
  \begin{center}
    \leavevmode
    \epsfxsize=10cm \epsfbox{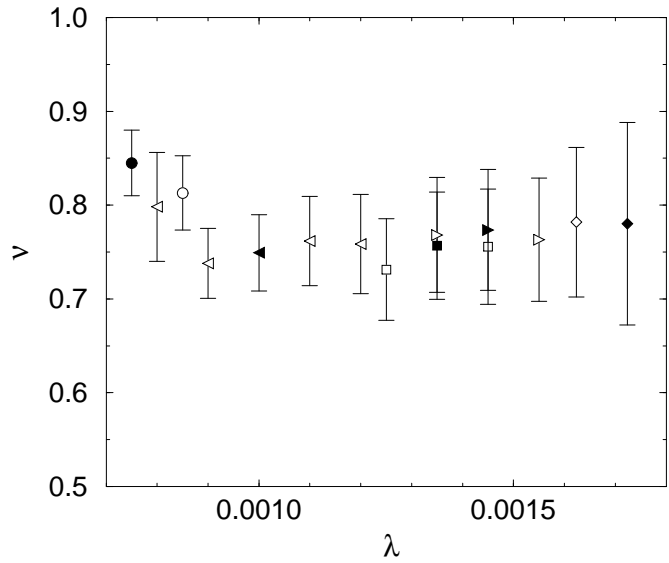}
    \caption{Exponent $\nu$ of finite size scaling of first Lee-Yang zero.}
    \label{fig:nu_ly0_6}
  \end{center}
\end{figure}

In Figure \ref{fig:nu_ly0_6} we show the exponent of scaling function
(\ref{eq:fss_ly0}).  The meanings of symbols are the same as in Figure
\ref{fig:res_ly0}.  For  $\lambda > \lambda_c$,  where there is no phase 
transition,  the exponent takes a value $\nu\approx 0.75$.   Below 
the endpoint $\lambda < \lambda_c$, the exponent shows some trend of 
increase, but not quite to the value $\nu=1$ expected for a first-order 
transition.  We think that this is due to insufficient volume sizes used 
in our simulation, for which corrections to the leading $1/V$ behavior are 
not negligible.  

\begin{figure}[p]
  \begin{center}
    \leavevmode
    \epsfxsize=10cm \epsfbox{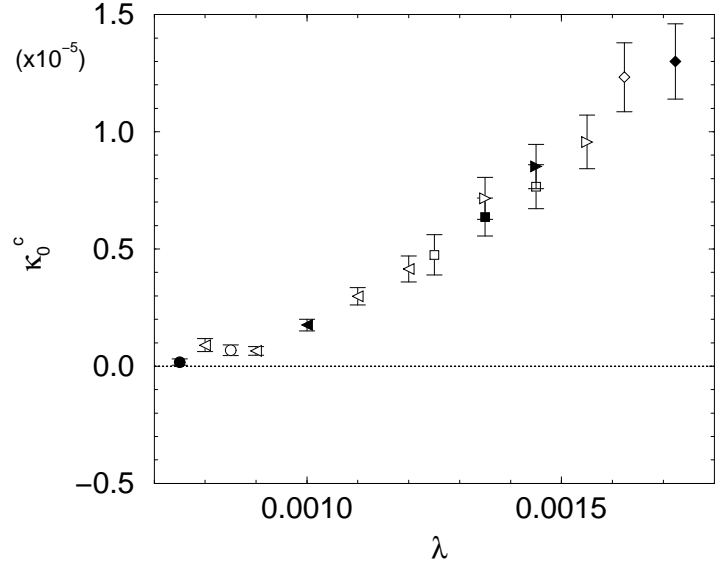}
    \caption{Same as in Fig.~\protect{\ref{fig:res_ly0}}. Quadratic
      polynomial is used for fit instead of power function.}
    \label{fig:res_ly0_Q_6}
  \end{center}
\end{figure}

To check this point we make an alternative fit of results for the first
zero adopting a quadratic ansatz in volume given by 
\begin{equation}
 \label{eq:fss_ly0_quad}
   {\rm Im}\kappa_0(V) = \kappa_0^c + C V^{-1} + D V^{-2}, 
\end{equation}
and show the results for $\kappa_0^c$ in Figure~\ref{fig:res_ly0_Q_6}.  
Clearly the infinite volume limit $\kappa_0^c$ starts to deviate from 
zero around $\lambda\approx 0.001$, which is consistent with the 
estimate of $\lambda_c$ above, albeit located at the lower end of 
the one standard deviation error band. 

We note that the quadratic ansatz (\ref{eq:fss_ly0_quad}), formally the 
first three terms of a Laurent series, is expected to be correct in case of 
a first-order phase transition, for which (\ref{eq:zeros}) describes the 
thermodynamic limit. However, it is not a valid 
assumption in the region of $\lambda$ where there is no phase transition. 
Therefore, unlike the case of Fig.~\ref{fig:res_ly0}, extrapolating 
the results of Fig.~\ref{fig:res_ly0_Q_6} from large to small values of 
$\lambda$ to 
estimate the location of the endpoint $\lambda_c$ is not justified.

\subsection{Binder Cumulant}
\label{subsec:bl}

Let us consider the Binder cumulant (cf. \cite{BINDER}) of the space-like 
link operator,
\begin{equation}
  \label{eq:binder}
  B_{L_s}(\kappa) \equiv 1
  \mbox{} - \frac{\langle L_s^4 \rangle}{3\langle L_s^2 \rangle^2};
  \; L_s = \frac{1}{3N_s^3N_t}\sum_{x,\mu=1,2,3}L_{x,\mu}
\end{equation}
The infinite volume limit of the minimum of this quantity should deviate 
from 2/3 for a first-order phase transition, 
while it should converge to 2/3 beyond the endpoint. 

We evaluate the minimum of the cumulant as a function of $\kappa$ 
for a given $\lambda$ and volume using  reweighting. 
We then use a scaling ansatz,
\begin{equation}
B_{L_s}^{\rm min} = B_{L_s}^c + C V^{-\nu},
\end{equation}
to extract the infinite-volume value $B_{L_s}^c$.

\begin{figure}[p]
  \begin{center}
    \leavevmode
    \epsfxsize=10cm \epsfbox{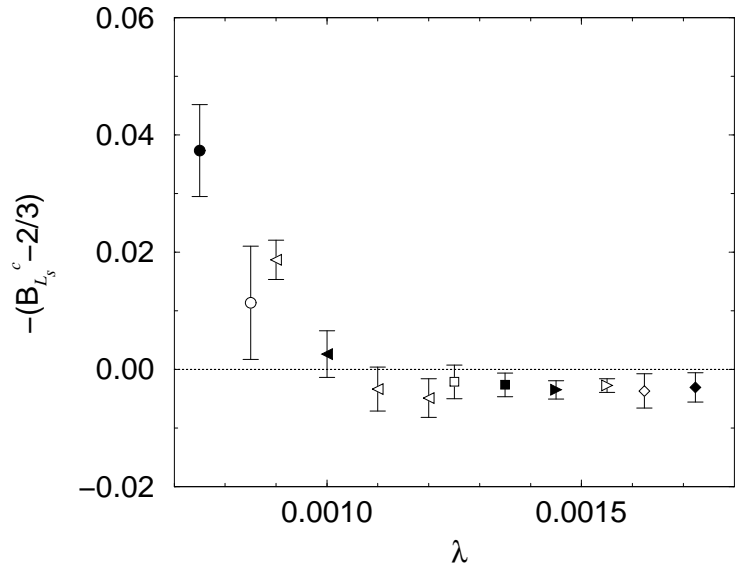}
    \caption{Minimum value of Binder cumulant of $L_s$
      at infinite volume limit as a function of $\lambda$.}
    \label{fig:res_bl_linkS_6}
  \end{center}
\end{figure}

In Fig.~\ref{fig:res_bl_linkS_6} we show $-(B_{L_s}^c - 2/3)$
as a function of $\lambda$, where the meanings of symbols are the same as 
in Fig.~\ref{fig:res_ly0}.
A change of behavior from non-vanishing values to those consistent with zero 
at $\lambda \approx 0.001$ shows that the first-order phase transition 
terminates around this value. 
Linearly extrapolating the two independent data at $\lambda = 0.00075$
and $0.001$ yields $\lambda_c = 0.00102(3)$ for the endpoint, 
which is consistent with the result (\ref{eq:endpoint}) 
from our study of Lee-Yang zeros. 
Note, however, that only two measured points are available for the linear 
extrapolation.  Therefore we can not make a statement on the goodness of the 
fit.  For this reason, we conservatively take the Lee-Yang 
value (\ref{eq:endpoint}) as our best estimate of the endpoint.

\section{Critical Higgs Boson Mass}
\label{sec:zeroT}

To determine the physical parameters characterizing the endpoint, namely the 
ratio of the Higgs boson mass to the W boson mass and  
the renormalized gauge coupling $g_R$, we have to perform zero-temperature 
simulations.
As in Refs. \cite{DESY94,DESY95,DESY96}, we extract the Higgs boson 
mass $m_H$ in lattice units from correlators of 
 $ \rho_x^2 \,$ and $L_{x, \mu}$.
The W boson mass in lattice units $m_W$ is obtained from the correlator of the
composite link fields
\begin{equation} \label{eq04}
  W_x  \equiv \sum_{r,k=1}^3 \frac{1}{2} {\rm Tr}
  (\tau_r \alpha_{x+\hat{k}}^+ U_{xk}\, \alpha_x ) \, ,
\end{equation}
where $\tau_r$ is the Pauli matrix and $\alpha_x$ is the angle part of $\Phi_x$ 
such that $\Phi_x \equiv\rho_x \alpha_x$ with $\alpha_x \in {\rm SU(2)}$.

Masses are extracted from the correlators fitting to a hyperbolic cosine 
plus a constant function.
Simple uncorrelated least-square fits
and correlated fits with eigenvalue smoothing proposed by 
Michael and McKerrell \cite{MICHAEL} are used.  
The application of this method is discussed in detail in Ref.~\cite{DESY96}.

The actual procedure of extracting the mass parameters is the
following.
First we determine the reasonable time intervals for fitting the 
correlator data.
The guideline is to choose as large an interval as possible with
reasonable $\chi^2$/d.o.f.\ value.
For this purpose correlated fits with eigenvalue smearing are 
used.  We find this to be necessary since the data are strongly 
correlated for different time distances.
Having fixed the fitting time interval, we next carry out uncorrelated 
fits.  To perform this fit, we divide the data sample into subsamples,  
and estimate the errors of correlators from the statistical fluctuations 
of subsample averages.  

The {\em best fit} value of the masses is taken to be the number
given by the uncorrelated fit.
The value of the Higgs boson mass is obtained by fitting to a linear 
combination of the two different correlators for $\rho_x^2$ and $L_{x,\mu}$. 
The errors on the masses are determined by jackknife analyses over subsamples.
The masses obtained by the correlated fits with eigenvalue smearing
are in all cases well within the error bars of the uncorrelated fits.

Our zero-temperature simulations are carried out 
at two points given by 
$(\lambda,\kappa=\kappa_c(\lambda,N_t=2))$ for $\lambda=0.0011$ and $0.00125$
employing several lattice sizes to examine finite-volume effects. 
The run parameters and results for masses are collected in Table
\ref{tab:param0}. The size of subsamples is typically 500 sweeps.

\begin{figure}[p]
\begin{center}
\leavevmode
\epsfxsize=10cm \epsfbox{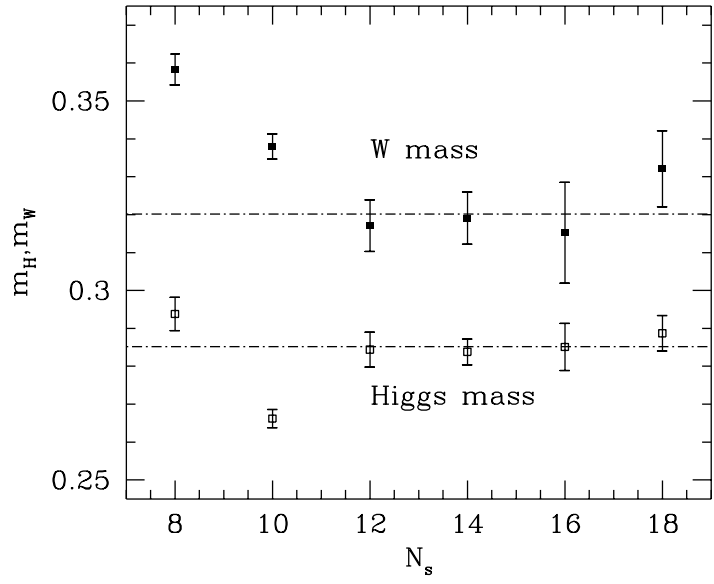}
\caption{Higgs and W masses in lattice units as a function of $N_s$ for 
$\lambda =0.0011$.}
\label{fig:m_11}
\end{center}
\end{figure}

Our results do not show significant volume dependence
(see Fig.~\ref{fig:m_11}), except for the two  
smallest spatial volumes $N_s=8^3$, $10^3$ for which somewhat different 
values are obtained compared to those of other volumes.  We then 
discard those results
and take an average over the rest of the volumes.  This yields
the values given in Table~\ref{tab:res0}.  Setting $M_W=80$ GeV, we obtain   
\begin{eqnarray}
M_H&=&70.9\pm 1.1 \mbox{GeV} \, (\lambda=0.0011) \\
M_H&=&76.8\pm 1.1 \mbox{GeV} \, (\lambda=0.00125).
\end{eqnarray}
Making a linear interpolation to the critical value 
$\lambda_c=0.00116(16)$ from the Lee-Yang zero analysis, 
we find 
\begin{equation}
M_{H,c}=73.3\pm 6.4 \mbox{GeV},
\end{equation}
where the error is dominated by that of $\lambda_c$.

From measurements of Wilson loops we also determine the values of the 
renormalized gauge coupling $g_R$ 
using the method described in Refs. \cite{DESY94,DESY95,DESY96}. 
The potential as a function of the distance $R$ is fitted by
\begin{equation} \label{pot}
V(R)= - \frac{A}{B} e^{-MR}  + C + D G(M,R,L_s) \, ,
\end{equation}
where $G(M,R,L_s)$ stands for lattice artifacts (cf. \cite{DESY95}). The 
potential is determined from the rectangular Wilson loops by fitting the 
time dependence with three exponentials. A stable fit is obtained in 
all cases. The potential is then fitted by (\ref{pot}) using all R values.
Our results for the fit parameters and $g_R^2$ for various spatial size 
lattices are shown in Table 
\ref{tab:g_R^2}. We see that $g_R$ is constant within errors.
The  averaged values are given in Table~\ref{tab:res0}. The values do 
agree within 
errors, showing that our simulations for the two $\lambda$ values correspond 
to the same renormalized gauge coupling. Therefore the linear extrapolation 
to $\lambda_c$ mentioned above is justified, since we use Higgs masses at 
equal renormalized gauge couplings.

Finally, let us try to estimate the effect of fermions and
the U(1) gauge boson on our result.  
We make this estimation through the perturbative expression for the 
parameter $x=\lambda_3/g_3^2$ of the dimensionally reduced model in terms 
of the physical parameters of the Standard Model \cite{Kajantie3}. 
Using our results for the Higgs boson mass and the renormalized gauge 
coupling, we find $x_c=0.121 \pm 0.020 $ for the endpoint. 
Including the effect of fermions and the U(1) gauge boson, 
this value corresponds to $M_{H,c}=80 \pm 7 $ GeV.
  
\section{Conclusions}
\label{sec:conc}

We have studied the endpoint of the finite-temperature first-order transition
of the SU(2) gauge-Higgs model on a space-time isotropic lattice of a 
temporal extension $N_t=2$.
The results from Lee-Yang zero and Binder cumulant analyses show 
that the first-order phase transition terminates at $\lambda_c=0.00116(16)$
and turns into a smooth crossover for $\lambda > \lambda_c$.

Setting $M_W =80$ GeV our result for the critical Higgs boson mass 
is $M_{H,c}=73.3\pm 6.4$ GeV. This is 
consistent within error with the value 
$M_{H,c}=74.6\pm0.9$ GeV \cite{Csikor982} 
obtained in a 4-dimensional anisotropic lattice simulation for the 
same temporal size.
The same work also reported that the critical mass decreases for larger 
temporal size, and extrapolates to $M_{H,c}=66.5\pm1.4$ GeV in the continuum
limit. This value is consistent with the 3-dimensional 
result $66.2$ GeV \cite{Leipzig97}.
Thus results from various methods, in three and four dimensions, 
agree well.

For a comparison with the experimental lower bound $M_H>87.9$ GeV
\cite{Aleph98} for the Higgs boson mass, we need to include the effect
of fermions and U(1) gauge boson.
The good agreement of critical mass from the four- and 
three-dimensional simulations noted above imply that this may be made 
perturbatively, with which we find $M_{H,c}=80 \pm 7 $ GeV for our 
$N_t=2$ simulation.  This value is about 10\% larger, albeit with a comparable 
error, than the result $M_{H,c}=72.4\pm1.7$ GeV in the continuum limit 
obtained from a 4-dimensional anisotropic study\cite{Csikor982}, possibly 
due to scaling violations.  
We also note that the 3-dimensional approach reported the 
values $M_{H,c}=72.4\pm0.9$ GeV\cite{Leipzig97} and
$M_{H,c}=72\pm2$ GeV\cite{Rumm98}. 
Combining all the available results, we conclude that the electroweak 
baryogenesis within the Minimal Standard Model is excluded. 

\section*{Acknowledgements}
Part of this work was carried out while Z.F. was visiting KEK by the Foreign 
Researcher Program of the Ministry of Education.  Part of numerical 
calculations was made on VPP-500/30 at the Information Processing 
Center of University of Tsukuba and the PMS-11G PC-farm in Budapest. 
This work is supported 
in part by Grants-in-Aid of the Ministry of Education of Japan 
(Nos.\ 09304029, 10640246), 
Hungarian Science Foundation grants (No.\ OTKA-T016240/T022929) and 
Hungarian Ministry of Education grant (No.\ FKP-0128/1997).



\begin{table}[p]
  \begin{center}
    \caption{Run parameters of finite temperature simulation
      and results of first Lee-Yang zero.
      Data used for analysis of susceptibility
      and Binder cumulant are marked with
      $\chi$ and $B$, respectively, in the last column.
      }
    \label{tab:paramT}
    \leavevmode
    \begin{tabular}{lllcclll}
      &&& \multicolumn{2}{c}{($\times 10^3$ sweep)} &&&
      \\
      \mbox{\hspace{20pt}}$\lambda$ & $N_s$ &
      \mbox{\hspace{16pt}}$\kappa$\mbox{\hspace{16pt}} &
      iteration  & bin size &
      \multicolumn{1}{c}{${\rm Re}\kappa_0$} &
      \multicolumn{1}{c}{${\rm Im}\kappa_0$} &
      \multicolumn{1}{c}{use}
      \\
    \hline
    0.00075
& 20 & 0.129114 & 100 & 2   & 0.1291133(23) & 0.0000477(20)     & $\chi,B$ \\
& 24 & 0.129103 & 100 & 2   & 0.1291068(12) & 0.0000285(11)     & $\chi,B$ \\
& 32 & 0.129102 & 100 & 4   & 0.12910273(91) & 0.00001351(45)   & $\chi,B$ \\
& 40 & 0.129100 & 100 & 6.25 & 0.12910086(72) & 0.00000762(26)  & $\chi,B$ \\
& 50 & 0.129100 & 120 & 10  & 0.12910041(51) & 0.00000411(17)   & $\chi,B$ \\
& 60 & 0.129100 & 180 & 10  & 0.129100308(303) & 0.000002321(51)& $\chi,B$ \\
    \\
    0.001
& 20 & 0.129340 & 100 & 2 & & & $B$ \\
& 20 & 0.129350 & 100 & 2 &
  \multicolumn{1}{l}{\raisebox{1.5ex}[0pt]{0.1293472(15)}} &
  \multicolumn{1}{l}{\raisebox{1.5ex}[0pt]{0.0000605(16)}} & $\chi$ \\
& 24 & 0.129330 & 100 & 2   & 0.1293357(18) & 0.0000432(18)    & $\chi,B$ \\
& 32 & 0.129328 & 100 & 2   & 0.12933093(122) & 0.00002136(75) & $\chi,B$ \\
& 40 & 0.129327 & 100 & 2.5 & 0.12932802(80) & 0.00001223(44)  & $\chi,B$ \\
& 50 & 0.129327 & 100 & 4   & & & $B$ \\
& 50 & 0.129328 & 100 & 4   &
  \multicolumn{1}{l}{\raisebox{1.5ex}[0pt]{0.12932797(37)}} &
  \multicolumn{1}{l}{\raisebox{1.5ex}[0pt]{0.00000763(26)}} & $\chi$ \\
& 60 & 0.1293275& 180 & 7.5 & 0.12932743(37) & 0.00000489(18) & $\chi,B$ \\
    \\
    0.00135
& 20 & 0.129660 & 100 & 1   & 0.1296888(34) & 0.0001167(36)    & $\chi,B$ \\
& 24 & 0.129650 & 100 & 1   & 0.1296619(29) & 0.0000819(32)    & $\chi,B$ \\
& 32 & 0.129644 & 100 & 1   & 0.1296465(20) & 0.0000542(20)    & $\chi,B$ \\
& 40 & 0.129640 & 100 & 2   & 0.1296426(15) & 0.0000293(11)    & $\chi,B$ \\
& 50 & 0.129639 & 120 & 2.5 & 0.12963782(137) & 0.00002016(88) & $\chi,B$ \\
& 60 & 0.129637 & 120 & 4   & 0.12963754(68) & 0.00001299(78)  & $\chi,B$ \\
    \\
    0.00145
& 20 & 0.129748 & 100 & 1   & 0.1297482(35) & 0.0000885(38)    & $\chi,B$ \\
& 24 & 0.129736 & 100 & 1   & 0.1297384(20) & 0.0000567(20)    & $\chi,B$ \\
& 32 & 0.129728 & 100 & 2   & 0.1297318(15) & 0.0000328(12)    & $\chi,B$ \\
& 40 & 0.129724 & 100 & 2   & 0.12972751(115) & 0.00002171(99) & $\chi,B$ \\
& 50 & 0.129722 & 120 & 2   & 0.12972654(80) & 0.00001529(79)  & $\chi,B$ \\
& 60 & 0.129724 & 120 & 4   & 0.12972517(61) & 0.00001146(79)  & $\chi,B$ \\
    \\
    0.0017235
& 20 & 0.129980 & 100 & 1   & & &\\
& 20 & 0.129990 & 100 & 1   &
  \multicolumn{1}{l}{\raisebox{1.5ex}[0pt]{0.1299875(20)}} &
  \multicolumn{1}{l}{\raisebox{1.5ex}[0pt]{0.0000951(19)}} & $\chi,B$  \\
& 24 & 0.129980 & 100 & 2.5 & 0.1299755(24) & 0.0000604(21) & $\chi,B$  \\
& 32 & 0.129966 & 100 & 1   & 0.1299654(15) & 0.0000383(12) & $\chi,B$  \\
& 40 & 0.129968 & 100 & 1   & 0.1299663(15) & 0.0000276(14) & $\chi,B$  \\
& 50 & 0.129965 & 100 & 2   & & & $\chi,B$ \\
& 50 & 0.129966 & 100 & 2   &
  \multicolumn{1}{l}{\raisebox{1.5ex}[0pt]{0.1299616(14)}} &
  \multicolumn{1}{l}{\raisebox{1.5ex}[0pt]{0.0000207(16)}} & \\
& 60 & 0.129962 & 120 & 4   & 0.12996122(71) & 0.00001585(74) & $\chi,B$ \\
    \end{tabular}
  \end{center}
\end{table}

\begin{table}[p]
  \begin{center}
    \caption{Run parameters of zero-temperature simulations and results
      for masses in lattice units.}
    \label{tab:param0}
    \leavevmode
    \begin{tabular}{lrcll}
      \multicolumn{1}{c}{$\lambda$} && ($\times 10^3$ sweep) &
      \\
      \multicolumn{1}{c}{$\kappa$} & $N_s^3 \times N_t$ &
      iteration  &
      \multicolumn{1}{c}{$m_H$} &
      \multicolumn{1}{c}{$m_W$} 
      \\
    \hline
    & $ 8^3 \times 20$ &  60 & 0.2938(44) & 0.3583(41)  \\
    & $10^3 \times 24$ &  75 & 0.2662(24) & 0.3380(33)  \\
    0.0011
    & $12^3 \times 28$ &  49 & 0.2844(46) & 0.3171(68)  \\
    0.129416
    & $14^3 \times 32$ &  34 & 0.2838(34) & 0.3191(69)  \\
    & $16^3 \times 36$ &  26 & 0.2851(62) & 0.3152(133) \\
    & $18^3 \times 36$ &  26 & 0.2887(47) & 0.3321(100) \\
    \hline
    & $ 8^3 \times 20$ &  60  & 0.2806(42)  & 0.3285(95)   \\
    & $10^3 \times 24$ &  75  & 0.2764(33)  & 0.3291(30)  \\
    0.00125
    & $12^3 \times 28$ &  49  & 0.2884(38) & 0.2992(51)  \\
    0.129532
    & $14^3 \times 32$ &  34  & 0.2851(56)  & 0.3037(58)  \\
    & $16^3 \times 36$ &  29  & 0.2863(91)  & 0.2941(64)  \\
    & $18^3 \times 36$ &  31.5& 0.2892(54)  & 0.2965(71)  \\
  \end{tabular}
\end{center}
\end{table}

\begin{table}[p]
  \begin{center}
    \caption{Averaged masses in lattice units and renormalized gauge couplings 
    from results in Table \protect\ref{tab:param0} excluding those for 
      the two smallest volumes.}
    \label{tab:res0}
    \leavevmode
    \begin{tabular}{lllll}
      \multicolumn{1}{c}{$\lambda$} &
      \multicolumn{1}{c}{$m_H$} &
      \multicolumn{1}{c}{$m_W$} &
      \multicolumn{1}{c}{$R_{HW}$} & 
      \multicolumn{1}{c}{$g_R^2$}
      \\
    \hline
    0.0011  & 0.2852(22) & 0.3202(41) & 0.8864(136) & 0.5712(27)\\
    0.00125 & 0.2877(26) & 0.2988(30) & 0.9607(134) & 0.5768(33)\\
  \end{tabular}
\end{center}
\end{table}

\begin{table}[p]
  \begin{center}
  \caption{Summary of the fit parameters for the static potential and the 
  renormalized gauge coupling.}
  \label{tab:g_R^2}
  \leavevmode
  \begin{tabular}{lrlllll}
  \multicolumn{1}{c}{$\lambda$} 
  \\
  \multicolumn{1}{c}{$\kappa$} & 
  \multicolumn{1}{c}{$N_s^3 \times N_t$} &
  \multicolumn{1}{c}{A} &
  \multicolumn{1}{c}{M} &
  \multicolumn{1}{c}{D} &
  \multicolumn{1}{c}{C} &
  \multicolumn{1}{c}{$g_R^2 \equiv \frac{16}{3}\pi A$} 
  \\
   \hline
  & $12^3 \times 28$ &0.03495(58) &0.3021(62) & 0.03941(68) &0.0968(2) & 0.5856(97) \\ 
  0.0011
  & $14^3 \times 32$ &0.03435(52) & 0.2783(90) & 0.03673(45) & 0.09672(21) &0.5755(87)  \\  
  0.129416
  & $16^3 \times 36$ & 0.03406(30) & 0.2898(107) & 0.03975(28) & 0.09632(13) &0.5707(50) \\
  & $18^3 \times 36$ & 0.03394(22) & 0.2791(42) & 0.04061(262) & 0.09633(3) & 0.5687(37) \\
  \hline
  & $12^3 \times 28$ & 0.03561(46) & 0.2788(121) & 0.02814(310) & 0.09751(31) &0.5966(77)  \\
  0.00125
  & $14^3 \times 32$ & 0.03456(70) & 0.2573(113) & 0.0353(57) & 0.09766(39) &0.5791(117)  \\
  0.129532
  & $16^3 \times 36$ & 0.03386(31) & 0.2559(79) & 0.0416(35) & 0.09740(12) & 0.5673(52) \\
  & $18^3 \times 36$ & 0.034442(35) & 0.2676(36) & 0.03831(41)& 0.09704(4) & 0.5770(59) \\
 \end{tabular}
  \end{center}
  \end{table}

\end{document}